\documentclass{article}
\usepackage{a4,latexsym,amsmath,epsf}
\newcommand{\bm}[1]{\mbox{\boldmath $#1$}}
\newcommand{\starprod}[2]{\renewcommand{\arraystretch}{0}
                          \begin{array}{c}\scriptstyle #2\\ 
                          \mbox{\Huge $\ast$}\\ \scriptstyle #1 
                          \end{array}\renewcommand{\arraystretch}{1}}
\begin{document}

\title{Aperiodic and correlated disorder in XY-chains: exact results}

\author{Joachim Hermisson \\
Institut f\"ur Theoretische Physik, Universit\"at
T\"ubingen,\\ Auf der Morgenstelle 14, 72076 T\"ubingen, Germany\\and\\
Institut f\"ur Theorie der kondensierten Materie,\\ Universit\"at Karlsruhe,
76128 Karlsruhe, Germany}
\maketitle
\begin{abstract}
We study thermodynamic properties, specific heat and susceptibility, of XY
quantum chains with coupling constants following arbitrary substitution 
sequences. Generalizing an exact renormalization group transformation, 
originally formulated for Ising quantum chains, we obtain exact relevance
criteria of Harris-Luck type for this class of models. For two-letter 
substitution rules, a detailed classification is given of sequences leading 
to irrelevant, marginal or relevant aperiodic modulations. We find that the 
relevance of the same aperiodic sequence of couplings in general will be 
different for XY and Ising quantum chains.
By our method, continuously varying critical exponents may be calculated 
exactly for arbitrary (two-letter) substitution rules with marginal 
aperiodicity. A number of examples are given, including the period-doubling, 
three-folding and precious mean chains. 
We also discuss extensions of the renormalization approach to a special class of
long-range correlated random chains, generated by random substitutions.
\end{abstract}

\section{Introduction}

Phase transitions and critical phenomena in Ising spin systems with 
(dis-)order of various nature (random, quasiperiodic, self-similar etc.) 
have been an active research area for many years.   
The main questions deal with the relevance of these kinds of
disorder to the thermodynamics of different models and the
characterization of new, disorder-induced universality classes. 
For random systems, the Harris criterion \cite{Harris} gives an
heuristic scaling argument for the relevance of disorder. This has later been
generalized by Luck \cite{L2} to general aperiodic disorder. The argument
is perturbative in nature and is expected to
hold for weak disorder: Comparing the local shift of the critical
point due to aperiodic modulations to the distance from criticality,
disorder should be {\em relevant} if the so-called {\em wandering
exponent} $\omega$ (measuring the geometrical fluctuations of the mean 
coupling constant) exceeds a critical value \cite{L2}
\begin{equation} \label{HL}
\omega_c = 1 - 1/(D\nu) \; .
\end{equation}
Here, $\nu$ is the correlation length exponent of the unperturbed
system and the model is disordered in $D$ co-ordinate directions.

Especially since the discovery of quasicrystals in 1984, the effect of
(deterministic) aperiodicity on the thermodynamical properties of
different models has been the topic of numerous, mostly numerical studies.
Universal behaviour was found for most quasicrystalline systems, like 
Ising models on Penrose and Amman-Beenker lattices \cite{GLO,ON1,SJR,S} and 
also in three dimensions \cite{ON2}. Marginal scaling, on the other hand, 
has been observed for the surface roughness of two-dimensional quasicrystals 
\cite{HeL,GL}. Compare also the review \cite{GB2} for further 
references. Although almost all results corroborate the Harris-Luck criterion, 
this is nevertheless somewhat more subtle for aperiodic or (more general)
correlated disorder than for randomly disordered systems. This is because it is 
in general not the fluctuations of the coupling constants directly, but of 
some related microscopic parameters (reduced coupling constants) that should 
be considered. While the statistical fluctuations of uncorrelated random 
couplings will normally lead to a similar fluctuational behaviour in the 
(a priori unknown) ``reduced couplings'', this is no longer guarantied if the coupling 
constants are correlated or even distributed according to a deterministic rule. The 
solvable Ising models on the Labyrinth \cite{BGB} provide an example where the 
strong correlations among the coupling constants due to the ``rapidity line 
parametrization'' enforces Onsager universal behaviour even for 
(according to \cite{L2}) relevant fluctuations of the mean coupling constant.

Analytical results have so far been obtained only for a small number of systems. 
With the exception of the Labyrinth models, which are, when solvable, somewhat 
non-generic in their aperiodicity, these are one-dimensional free-fermion models,
like tight-binding models or quantum chains. Most results rely moreover on a 
special choice of the aperiodic orderings (like the Fibonacci model) which 
makes them applicable to efficient trace-map methods or renormalization 
techniques derived therefrom \cite{KKT,AS}. 
Independently of trace map properties, the surface magnetization
of aperiodic {\em Ising quantum chains} with constant transverse field 
has been calculated exactly for certain substitution sequences \cite{IT2}. 
Only recently, a decimation procedure in real space has been introduced
\cite{IT} (again for particular substitution rules) that later could be
generalized to obtain analytically the scaling properties of the whole class
of Ising quantum chains with coupling constants following arbitrary 
substitution rules \cite{HGB}. This lead to an analytical confirmation
of Luck's relevance criterion for these models.

In this article, after an introduction of the model in section 2, we show in section 3
how the renormalization approach, as formulated for
the Ising quantum chains, can be extended to aperiodic XY spin
chains. It turns out that, for a given sequence of couplings, the influence 
of the induced disorder may be different in the two models. Nevertheless,
fluctuations turn out to be the basic concept for the demarcation of
relevant from irrelevant disorder. However, the fluctuations of the sequence
of coupling constants and of the induced sequence of reduced coupling 
constants, that determine the critical behaviour here, behave -- in contrast 
to the randomly disordered case -- in general differently for aperiodic order.
This way, the Harris-Luck relevance criterion may be adapted to XY
spin chains or, equivalently, to tight-binding models with aperiodic
hopping. We calculate the scaling exponent of the central spectral gaps at 
criticality and derive therefrom in section 4 (following \cite{LN}) the critical 
scaling behaviour of the specific heat and the zero-field susceptibility. Connections
to localization properties of tight binding models are briefly mentioned.
In section 5, we show how known results from trace-map approaches can be
rederived, clarifying their origin in this broader context. As
examples, we also give some new scaling exponents for different
aperiodic chains with marginal disorder. In section 6, an extension of the 
renormalization approach to random substitution rules is proposed; finally we
conclude with a short discussion.

\section{The model}

The system we are concerned with here is defined by the following 
quantum Hamiltonian:
\begin{equation}
H_{N}^{} = \sum_{j=1}^{N}\left(
     \varepsilon_{j}^{x}\sigma^{x}_{j}\sigma^{x}_{j+1} + 
     \varepsilon_{j}^{y}\sigma^{y}_{j}\sigma^{y}_{j+1}\right) \; .
\label{eq:QC}
\end{equation}
The coupling constants $\varepsilon_{j}^{x,y}$ are 
site-dependent and the operators $\sigma^{x,y}_{j}$ denote 
Pauli's matrices acting on the $j$\/th site. Boundary conditions may be 
chosen periodical ($\sigma_{N+1} = \sigma_1$) or free ($\varepsilon_N =0$). 

For a general set of coupling constants, this model is equivalent with
a free fermion field \cite{Lieb, Smith}, the fermionic excitation
energies $\Lambda_q$ satisfying the linear difference equations
\begin{eqnarray}\label{diff1}
\Lambda_q \psi_j^{(q)} &=& \varepsilon_{j-1}^{x} \phi_{j-1}^{(q)}
+\varepsilon_{j}^{y} \phi_{j+1}^{(q)} \\ \label{diff2}
\Lambda_q \phi_j^{(q)} &=& \varepsilon_{j-1}^{y} \psi_{j-1}^{(q)}
+\varepsilon_{j}^{x} \psi_{j+1}^{(q)} \; .
\end{eqnarray}
Defining
\begin{eqnarray}
\eta_{2j}^{(q)} = \phi_{2j}^{(q)} && \eta_{2j-1}^{(q)} =
\psi_{2j-1}^{(q)} \\
\hat{\eta}_{2j}^{(q)} = \psi_{2j}^{(q)} && \hat{\eta}_{2j-1}^{(q)} =
\phi_{2j-1}^{(q)} \; ,
\end{eqnarray}
these equations decouple into the eigenvalue problems of two independent tight
binding models with aperiodic hopping. 
This decoupling can also be carried through on the level of the 
spin chain Hamiltonian itself and has been used there to analyse 
XY-chains with random bonds \cite{Fisher}. 
Difference operators of the kind (\ref{diff1}, \ref{diff2}) underly various physical models
and may also be interpreted as a phononic model with varying spring
constants or the transition matrix of a one-dimensional random walk in
an aperiodic environment. 
The Ising quantum chain with transverse magnetic field in its
fermionic form also gives rise to a similar set of equations, the 
field variables replacing the $\varepsilon^y$ couplings. 
In \cite{HGB}, a renormalization scheme has been defined for the case
of a {\em uniform} magnetic field (or, more generally, field variables
depending on the neighbouring coupling constants), thereby effectively decoupling
the degrees of freedom that finally enter the renormalization scheme.
In our situation, however, the $\varepsilon^y$ couplings will not be determined
through their neighbourhood, but, together with the $\varepsilon^x$
couplings, follow the aperiodic sequence that defines the model.  

For uniform or randomly distributed coupling constants, the XY- chain 
exhibits a zero temperature phase transition from an X- to an 
Y-ferromagnetically ordered phase at $[\ln \varepsilon^x]_{av} 
= [\ln\varepsilon^y]_{av}$.
For the quasiperiodic Fibonacci sequence, a non-universal scaling law has been
found for isotropic couplings ($\varepsilon_i^x = \varepsilon_i^y$)
with scaling exponents depending on the coupling constants \cite{LN}.

Here, the site-dependent coupling constants $\varepsilon_{j}^{x,y}$ shall be
chosen according to a sequence generated by an arbitrary substitution rule
\begin{equation}
\varrho: a_i \;\rightarrow\; w_i 
\end{equation}
with words $w_i$ taken from an $n$-letter alphabet ${\cal A}$ with
letters $a_1,\ldots,a_n$. 
In the following, $w_i^\ell$ shall denote the $\ell$th letter and
$\#_\alpha(w_i)$ the total number of letters $\alpha$ in $w_i$. 
Since also (non-overlapping) pairs of consecuting letters in a word 
will play an important role, we further define $\#_{\alpha\beta}(w_i)$ to
give the number of pairs $(\alpha\beta)$ contained in $w_i$. Finally, 
let $|w_i|$ be the length of $w_i$.

The parameters of the aperiodic model are given by the ratios of
the $\varepsilon_{a_i}^{x,y}$'s. We define, on a logarithmic scale,
\begin{equation}
r_{ij} \equiv \ln \varepsilon_{a_i}^x + \ln \varepsilon_{a_i}^y - 
\ln \varepsilon_{a_j}^x - \ln \varepsilon_{a_j}^y
\end{equation}
and
\begin{equation}
\Delta_i \equiv \ln \varepsilon_{a_i}^x - \ln \varepsilon_{a_i}^y \;.
\end{equation}
The $n$-letter model thus contains $n-1$ independent variables
which parametrize the strength of {\em isotropic aperiodicity} and $n$
parameters $\Delta_i$ which determine the {\em aperiodic anisotropy} of
the model.
For notational clarity, we will restrict discussions from now on 
to substitution systems on a two-letter alphabet, and comment only briefly on
extensions to general $n$-letter substitutions, which in most cases
can be dealt with along the same lines. We assume the following normal form 
for the substitution rule
\begin{equation} \label{subs}
\varrho:
\begin{array}{rcl}
a & \rightarrow & w_a \; \equiv aw'_a  \\
b & \rightarrow & w_b \; \equiv bw'_b
\end{array} \;,
\end{equation}
which does not mean a restriction as any two-letter substitution rule may
be transformed accordingly without changing the model \cite{HGB}.
In principle, the special form for $\varrho$ is not needed to make 
our renormalization group work, but it simplifies some of the calculations.
The two-letter model contains three independent parameters 
\begin{equation}
r \equiv \ln \frac{\varepsilon_a^x\varepsilon_a^y}{\varepsilon_b^x\varepsilon_b^y}
\end{equation} 
and
\begin{equation}
\bm{\Delta} \equiv \begin{pmatrix} \Delta_{a}\\ \Delta_b \end{pmatrix} = 
\begin{pmatrix} \ln (\varepsilon_{a}^x/\varepsilon_{a}^y) \\
\ln (\varepsilon_{b}^x/\varepsilon_{b}^y) \end{pmatrix} \;.
\end{equation}
Some basic properties of the sequence generated by $\varrho$
are already contained in the corresponding substitution matrix
\begin{equation}
\bm{M}_\varrho := 
\left(\begin{array}{@{\,}r@{\;\;}r@{\,}}
\#_a(w_a)&\#_a(w_b)\\ \#_b(w_a)&\#_b(w_b) \end{array}\right)
\end{equation}
with eigenvalues $\lambda_\pm$.
The leading eigenvalue, $\lambda_+$, gives the asymptotic scaling
factor of the system size with the number of iterated substitutions,
the entries of the corresponding (statistically normalized)
eigenvector $\bm{v}_+$ determine the frequencies $p_{a,b}$ of the letters
$a,b$ in the limit word \cite{Q}. The further eigenvalues of the substitution
matrix are connected to fluctuation modes present in the
sequence. Especially, the next to leading eigenvalue (here
$\lambda_-$, of course) determines the fluctuation (or wandering)
exponent \cite{L1}
\begin{equation} \label{omega}
\omega_\varrho = \frac{\ln|\Lambda_-|}{\ln \lambda_+} \; .
\end{equation}
For two-letter substitution rules in particular, a
classification of the substitution matrices according to the fixed
points of their {\em reductions modulo 2} will be helpful. Up to the
exchange of the letters $a$ and $b$, there are five possible fixed
points of $[\bm{M}_\varrho]_{\mbox{\footnotesize mod}\, 2}$, grouping into 
three cases, for which we will show the following:
\begin{enumerate}
\item
\begin{align}
\text{a)}\qquad&
[\bm{M}_\varrho]_{\mbox{\footnotesize mod}\, 2}
=\left(\begin{array}{@{\,}r@{\;\;}r@{\,}} 
0&0\\0&0\end{array}\right) \qquad \left[ =
\left(\begin{array}{@{\,}r@{\;\;}r@{\,}}
1&1\\1&1\end{array}\right)^2  = \left(\begin{array}{@{\,}r@{\;\;}r@{\,}} 
0&1\\0&0\end{array}\right)^2 \right]
\\
\text{b)}\qquad&
[\bm{M}_\varrho]_{\mbox{\footnotesize mod}\, 2} 
= \left(\begin{array}{@{\,}r@{\;\;}r@{\,}} 
1&0\\1&0\end{array}\right)
\end{align}
The chain is in its critical phase at $T=0$ iff
\begin{equation}\label{crit1}
\bm{v}_+ \cdot \bm{\Delta} = \pm \frac{r}{\lambda_+} \Big[
p_a (\#_{ab} - \#_{ba})(w_a) + p_b(\#_{ab}-\#_{ba})(w_b) \Big] \;.
\end{equation}
This condition holds in particular if $(\Delta_a \#_a + \Delta_b \#_b)(w_\alpha)
 = \pm r (\#_{ab} - \#_{ba})(w_\alpha)$ where we find unperturbed 
scaling behaviour $(z = 1)$. Otherwise, the relevance of the aperiodicity to 
the critical scaling is identical to the one for Ising quantum chains 
with couplings following the same aperiodic sequence: We obtain $z = 1$, 
marginal scaling with 
non-universal $(1<z<\infty)$, or random-like behaviour $(z = \infty)$ for 
$|\lambda_-|$ lower, equal, or larger than $1$ respectively.
Finally, off criticality, there is a gap in the fermionic integrated density 
of states (IDOS) at the band center.
\item
\begin{align}
\text{a)}\qquad&
[\bm{M}_\varrho]_{\mbox{\footnotesize mod}\, 2} 
= \left(\begin{array}{@{\,}r@{\;\;}r@{\,}} 
1&1\\0&0\end{array}\right)\\
\text{b)}\qquad&
[\bm{M}_\varrho]_{\mbox{\footnotesize mod}\, 2} 
= \left(\begin{array}{@{\,}r@{\;\;}r@{\,}} 
1&0\\0&1\end{array}\right)\qquad \left[ = 
\left(\begin{array}{@{\,}r@{\;\;}r@{\,}} 
0&1\\1&0\end{array}\right)^2  = \left(\begin{array}{@{\,}r@{\;\;}r@{\,}} 
1&1\\1&0\end{array}\right)^3 
= \left(\begin{array}{@{\,}r@{\;\;}r@{\,}} 
1&0\\1&1\end{array}\right)^2 \right]
\end{align}
We obtain a critical ground state and a vanishing gap in the center of the IDOS
iff
\begin{equation}\label{crit2}
\bm{v}_+ \cdot \bm{\Delta} = 0 \; .
\end{equation}
In particular, this implies that the model is critical for any kind of 
{\em isotropic aperiodicity} $(\Delta_a = \Delta_b =0)$. In this case, the 
relevance of the aperiodic modulation is determined by the renormalization 
group eigenvalue $\lambda_{xx} = |\#_{ab} - \#_{ba}|(w_aw_b)$ and is completely 
independent of the relevance of the same aperiodic ordering in the couplings of 
the corresponding Ising quantum chain. The relevance of 
{\em aperiodic anisotropy}, on the other hand, again agrees with the Ising case.
If isotropic and anisotropic aperiodicity are both marginal, we obtain a 
nonuniversal scaling exponent $1<z<\infty$ depending on two parameters.
\item
\begin{equation} \label{case5}
[\bm{M}_\varrho]_{\mbox{\footnotesize mod}\, 2} 
= \left(\begin{array}{@{\,}r@{\;\;}r@{\,}} 
1&0\\0&0\end{array}\right)
\end{equation}
The criticality condition is again given by (\ref{crit2}).
Isotropic aperiodicity is irrelevant if $\#_{ab}(w_b)=\#_{ba}(w_b)$ {\em or} 
if $2\#_{b0}(w_b) = \#_{b}(w_b)$. Here, $\#_{b0}(w_b)$ gives the number of 
$b$'s in $w_b$ with $w_b = w_1bw_2$ and $\#_a(w_1)$ even. 
In all other cases, the induced disorder is relevant and $z=\infty$.
Again, the relevance of {\em aperiodic anisotropy} agrees with the Ising case. 
\end{enumerate}
Please keep in mind that discussions may be concentrated on the 
five fixed points, since any
substitution rule can be transformed into an appropriate one by taking a 
suitable power. This clearly does not change the limit chain.
In the following, we show how to derive these relevance criteria within an exact
renormalization scheme.

\section{The renormalization group}

In this section, we generalize the renormalization procedure introduced for the
Ising quantum chain in \cite{HGB}. It relies on a decimation process found 
in \cite{IT} for particular substitution rules and uses a special 
{\em star-product technique}, originally developed many years ago in the 
context of scattering theory (see \cite{RED}). 

We introduce ${\cal S}$ transfer matrices for the {\em decoupled} sets 
of difference equations (\ref{diff1}, \ref{diff2})
\begin{equation}
\begin{pmatrix} \eta_{2k-1}^{}\\ \eta_{2l}^{} \end{pmatrix} = {\cal S}_{k|l} 
\begin{pmatrix} \eta_{2k}^{}\\ \eta_{2l-1}^{} \end{pmatrix} \; ; \quad 
\begin{pmatrix} \hat{\eta}_{2k-1}^{}\\ \hat{\eta}_{2l}^{} \end{pmatrix} 
= {\cal \hat{S}}_{k|l} 
\begin{pmatrix} \hat{\eta}_{2k}^{}\\ \hat{\eta}_{2l-1}^{} \end{pmatrix}\;.
\end{equation}
Since both systems are related under the exchange of $\varepsilon^x$ and 
$\varepsilon^y$ couplings, we may concentrate on the $\eta$ equations from now on. 
The ${\cal S}$ matrices transform by $\ast$-products like
\begin{equation}
{\cal S}_{k|l} = {\cal S}_{k|k+1} \ast {\cal S}_{k+1|k+2} \ast \ldots 
{\cal S}_{l-1|l} \equiv \starprod{i=1}{l-k} {\cal S}_{k+i-1|k+i}
\end{equation}
where $k<l$ and the $\ast$-product of two matrices is defined as
\begin{equation}\label{star}
\begin{pmatrix} a_1&b_1 \\ c_1&d_1\end{pmatrix} \ast 
\begin{pmatrix} a_2&b_2 \\ c_2&d_2\end{pmatrix} =
\begin{pmatrix} a_1&0\\0&d_2\end{pmatrix} + \frac{1}{1- d_1 a_2} 
\begin{pmatrix} b_1 c_1 a_2 & b_1 b_2 \\ c_1 c_2 & d_1 b_2 c_2 \end{pmatrix}\;.
\end{equation}
The form of the elementary matrices follows from (\ref{diff1}, \ref{diff2}). For 
$\varepsilon_{2k-1}^y = \varepsilon_\alpha^y$, 
$\varepsilon_{2k}^x = \varepsilon_\beta^x$ and 
$\varepsilon_{2k+1}^y = \varepsilon_\gamma^y$ we obtain
\begin{equation}\label{smatrix}
{\cal S}_{k|k+1} \equiv
{\cal S}_{\alpha\beta\gamma} = \begin{pmatrix} 
\Lambda / \varepsilon_\alpha^y 
& -\varepsilon_\beta^x/\varepsilon_\alpha^y \\
- \varepsilon_\beta^x/\varepsilon_\gamma^y 
& \Lambda / \varepsilon_\gamma^y
\end{pmatrix}\;.
\end{equation}
In the corresponding problem of the aperiodic Ising quantum chain with constant
transverse field $h=1$, each ${\cal S}$ matrix contains just one coupling
constant ($\varepsilon_\beta^x$ in (\ref{smatrix}), setting all
$\varepsilon^y \to h = 1$), enabeling an easy renormalization procedure 
reversing the substitution steps by $\ast$ multiplication of the corresponding
${\cal S}$ transfer matrices. With an ${\cal S}$ matrix depending on three 
consecuting coupling constants, the substitution rule has to be redefined 
appropriately in order to make an inverse transformation possible by taking 
${\ast}$ products. As the part of a coupling constant in the Ising model is
just taken by the ratio of two consecuting coupling constants in the XY case,
a substitution rule on pairs of letters (actually rather than triples) is needed 
here to make the renormalization procedure work. The Ising problem on $n$ 
letters thus corresponds to a problem of dimension $n^2$ in the XY case -- 
if such a substitution rule can be found at all\footnote{The pair substitution 
needed here is entirely different from the one used to describe the Ising 
quantum chain with coupling constants depending on the two end-points 
(site-problem), considered in \cite{Q,TBB}. In the Ising site problem, the chain
is divided into overlapping pairs and each coupling constant appears in two pairs
(with each of its neighbours), but only in one pair here. This may lead to a 
much more pronounced change in the scaling behaviour, see below.}. For the 
moment, let us concentrate on the cases 1 and 2 in the above classification,
where a substitution rule of the desired form is easily constructed.
Here, $|w_\alpha|+|w_\beta|$ is even for any $\alpha,\beta \in \{a,b\}$ and 
we obtain a pair substitution 
\begin{equation} \label{rho2}
\varrho_2 : (\alpha\beta) \to w_{\alpha\beta} \equiv w_\alpha w_\beta 
\end{equation}
with substitution matrix
\begin{equation}\label{pairsub}
\bm{M}_2 = 
\left(\begin{array}{@{\,}r@{\;\;}r@{\;\;}r@{\;\;}r@{\,}} 
\#_{aa}(w_{aa})&\#_{aa}(w_{ab})&\#_{aa}(w_{ba})&\#_{aa}(w_{bb})\\
\#_{ab}(w_{aa})&\#_{ab}(w_{ab})&\#_{ab}(w_{ba})&\#_{ab}(w_{bb})\\
\#_{ba}(w_{aa})&\#_{ba}(w_{ab})&\#_{ba}(w_{ba})&\#_{ba}(w_{bb})\\
\#_{bb}(w_{aa})&\#_{bb}(w_{ab})&\#_{bb}(w_{ba})&\#_{bb}(w_{bb})
\end{array}\right) \;.
\end{equation}
Let $\lambda_i$ be the eigenvalues of $\bm{M}_2$, 
$\lambda_1 = \lambda_{PF}$ being the Perron-Frobenius one. 
We now may adjoin each transfer matrix ${\cal S}_{\alpha\beta\gamma}$ 
to the pair $(\alpha\beta)$ and define corresponding reduced coupling constants
\begin{equation} \label{redcoup}
\mu_{\alpha\beta} \equiv \ln \varepsilon_\beta^x - \ln \varepsilon_\alpha^y \;.
\end{equation}
The renormalization transformation is now obtained by reversing the substitution
procedure (\ref{rho2})
\begin{align} \label{sren}
\tilde{\cal S}_{\alpha\beta\gamma} &\equiv 
\begin{pmatrix} 
\tilde{\kappa}_{\alpha\beta}^+ \tilde{\Lambda} / \varepsilon_\alpha^y 
& \pm \exp{\tilde{\mu}_{\alpha\beta}} \\
\pm (\varepsilon_\alpha^y/\varepsilon_\gamma^y)\exp{\tilde{\mu}_{\alpha\beta}} 
& \tilde{\kappa}_{\alpha\beta}^- \tilde{\Lambda} / \varepsilon_\gamma^y
\end{pmatrix} 
\\ &\equiv 
\starprod{i=1}{|w_{\alpha\beta}|/2} {\cal S}_{w_{\alpha\beta}^{2i-1} 
w_{\alpha\beta}^{2i} w_{\alpha\beta}^{2i+1}}
\end{align} 
where $w_{\alpha\beta}^{|w_{\alpha\beta}|+1} \equiv \gamma$.
In (\ref{sren}), we have introduced weights $\kappa_{\alpha\beta}^\pm$ 
as additional parameters in our renormalization group. These account for the 
fact that the renormalization blocks in general will be asymmetric 
(which causes $\kappa^+$ and $\kappa^-$ to differ), may contain 
different coupling constants and also vary in length, which results 
in different local weights $\kappa_{\alpha\beta}$ and $\kappa_{\alpha'\beta'}$.
Note that formally, in (\ref{sren}), only the $\varepsilon^x$ couplings are 
renormalized, while the $\varepsilon^y$'s just keep their values. 

The renormalization flow on the critical surface (defined through the 
vanishing of the ``mass gap'', $\Lambda \equiv 0$) may now be directly 
obtained from (\ref{star}) as
\begin{equation}
\bm{\tilde{\mu}} = \bm{M}_2^t \bm{\mu}
\end{equation}
where, in the 2-letter case,
\begin{equation}
\bm{\mu} = \begin{pmatrix} \mu_{aa}\\ \mu_{ab} \\ \mu_{ba} \\ \mu_{bb} 
\end{pmatrix} = \frac{1}{2}
\begin{pmatrix} 2\Delta_a \\ \Delta_a + \Delta_b + r \\ 
\Delta_a + \Delta_b-r \\ 2 \Delta_b \end{pmatrix}  \;. 
\end{equation}
The reduced coupling constants thus transform
with the  {\em transpose of the pair substitution matrix} $\bm{M}_2$.
Note that the RG transformations are the same in both decoupled eigenvalue
systems, however with different initial conditions 
$(\Delta_{a,b} \to - \Delta_{a,b})$ of the reduced couplings.
Obviously, the ``Onsager fixed-point'' of the uniform model just corresponds
to $\bm{\mu} \equiv \bm{0}$ and the eigenvalues and eigenvectors of $\bm{M}_2^t$
directly reveal the {\em RG eigenvalues} and {\em scaling fields} of the model. 
The contributions in the directions of different scaling fields are measured
by the scalar products of $\bm{\mu}$ with the eigenvectors $\bm{V_i}$ 
of $\bm{M}_2$. In contrast to the renormalization of the Ising quantum chains 
the vector of reduced couplings $\bm{\mu}$ is constrained here to a $2n-1$ 
dimensional subspace of the vector space spanned by the frequencies of $n^2$ 
different pairs of $n$ letters. It may thus happen that certain scaling fields 
vanish for arbitrary choice of $\bm{\mu}$ which indicates that the corresponding 
fluctuational mode is not present in the problem. The contribution to the      
leading scaling field (with corresponding eigenvalue $\lambda_{PF} = \lambda_+$, 
see below), however, does not vanish for a generical $\bm{\mu}$, but
\begin{equation} \label{critcond}
 \bm{\mu} \cdot \bm{V}_{PF} = \sum_{(\alpha\beta)} p_{(\alpha\beta)} 
\mu_{\alpha\beta} = [\ln \varepsilon^x]_{av} - [\ln \varepsilon^y]_{av} = 0
\end{equation}
leads to the well-known {\em criticality condition} for these models (e.g.\/ 
\cite{Pfeuty}). Any non-zero contribution immediately drives the system 
off the critical surface ($\bm{\mu} \to \pm \infty$) into the X or Y 
ferromagnetic phase. Note, however, that the whole model is critical if the 
criticality condition is fulfilled for just one of the decoupled subsystems.  

The presence of aperiodic disorder in the chain leads to non-zero 
contributions in the direction of the additional scaling fields. Let $\lambda_2$ be
the largest eigenvalue with a non-vanishing scaling field for critical couplings.
If this scaling field is relevant, that is, iff the {\em wandering exponent} of
the sequence of reduced couplings is positive,
\begin{equation} \label{omega2}
\omega_2 \equiv \frac{\ln |\lambda_2|}{\ln \lambda_{1}} > 0\,,
\end{equation}
the system flows to the corresponding {\em strong coupling fixed point} 
of the RG, where the reduced couplings divide into two types, taking the values 
$\pm \infty$ respectively. 
A (simple) eigenvalue $|\lambda_2| = 1$ leads to a marginal 
scaling field and the system flows to a fixed line with continuously varying
exponents. This confirmes the Harris-Luck criterion for these models, since
$D = \nu = 1$ for XY and Ising quantum chains, and thus $\omega_c = 0$ according to 
(\ref{HL}).

Before we discuss the effects of marginal or relevant aperiodicity to the 
critical behaviour, let us have a closer look at the spectra of the pair 
substitution matrices and the induced RG flows in the cases 1 and 2 above.
It is worthwhile to consider for a moment the situation of {\em vanishing 
isotropic aperiodicity}, $r=0$. Using
\begin{equation}
2\#_{aa}(w_{\alpha\beta}) + \#_{ab}(w_{\alpha\beta}) + 
\#_{ba}(w_{\alpha\beta}) = \#_a(w_\alpha) + \#_a(w_\beta)
\end{equation}
we recognize that the vector of the anisotropy parameters $\bm{\Delta}$ 
transforms with the transpose of the original substitution matrix 
$\bm{M}_\varrho$
\begin{equation} \label{anisotran}
\tilde{\bm{\Delta}} = \bm{M}_\varrho^t \bm{\Delta} \;.
\end{equation}  
However, this means that the fixed point structure and the RG flow near the 
fixed points is identical for XY-chains with aperiodic anisotropy and the 
aperiodic Ising spin chain analysed previously \cite{HGB}. 
Of course, (\ref{anisotran}) implies that the spectrum of $\bm{M}_\varrho$
is contained in the one of  $\bm{M}_2$. These properties generalize without
any change to $n$ letter substitutions with $|w_i|$ all even or all odd.
Let us now see what happens to the RG flow when isotropic aperiodicity
is turned on. 
\begin{itemize}
\item
In case 1, where both $|w_a|$ and $|w_b|$ are even at the fixed point 
of $[\bm{M}_\varrho]_{\mbox{\footnotesize mod}\, 2}$, the entries of $\bm{M}_2$ are
related as
\begin{equation}
\#_{\alpha\beta}(w_{\alpha'\beta'}) = \#_{\alpha\beta}(w_{\alpha'})
+\#_{\alpha\beta}(w_{\beta'})
\;;\;\alpha,\beta,\alpha',\beta'\in\{a,b\}
\end{equation}
and we conclude that the spectrum is just the set 
$\{\lambda_+,\lambda_-, 0, 0\}$. 
Using $\lambda_+ p_{ab} = p_a \#_{ab}(w_a)+ p_b \#_{ab}(w_b)$ we recognize 
that the criticality condition (\ref{critcond}) reduces to equation (\ref{crit1}) 
in this case. Clearly, for substitution rules with $\#_{\alpha\beta}(w_\gamma) = 
\#_{\beta\alpha}(w_\gamma) \,\forall \,\alpha,\beta,\gamma$, isotropic 
aperiodicity is completely irrelevant. In any other case, 
isotropic aperiodicity and aperiodic anisotropy are coupled in (\ref{crit1}). 
The critical manifold (which always has to be symmetric under the exchange of 
indices $x$ and $y$) splits into two submanifolds corresponding to the 
criticality of the decoupled subsystems which intersect at $r=0$. 
Put in other words, isotropic aperiodicity deforms (and splits) the
critical manifold, leading to a finite renormalization of the 
anisotropy parameters in the first RG step, but fixed point structure 
and RG flows remain unchanged otherwise. The changes of the critical 
behaviour induced by modulated aperiodicity remain closely related to the Ising case. 
This scenario generalizes without change to the $n$-letter case. 
\item
In case 2, $|w_a|$ and $|w_b|$ are both odd and we establish the 
following relations
\begin{gather}
\#_{ab}(w_{aa}) = \#_{ba}(w_{aa}); \quad \#_{ab}(w_{bb})=\#_{ba}(w_{bb}) \\
\#_{ab}(w_{ab})-\#_{ba}(w_{ab})=\#_{ba}(w_{ba})-\#_{ab}(w_{ba})
\end{gather}
{\em In absence of anisotropy} this means that 
$\bm{\mu}$ is already an eigenvector of $\bm{M}_2^t$ with eigenvalue 
$\lambda_{xx} = (\#_{ab}-\#_{ba})(w_{ab})$, 
independent of the detailed form of the substitution rule. Note that 
$\lambda_{xx} \le \lambda_+$, and $\lambda_{xx} = \lambda_+$ only in
the degenerated case where $w_{ab} = (ab)^n$ and the limit chain is periodic 
(with period $ab$). For all other cases, we conclude that the XX chain is 
critical for any amount of aperiodicity induced by substitution rules of the
given form. In the XY case, the parameters $r$ and $\bm{\Delta}$ 
renormalize independently. In other words, the isotropic aperiodicity
leaves the anisotropy parameters unrenormalized and does not deform the critical
surface, but introduces an additional scaling field in the RG. Since
this may be marginal or relevant, the scaling behaviour in general will be independent
of the ``Ising case''. Note that the remaining eigenvalue $\lambda_4 = 
\delta_{w_a^{|w_a|},a} + \delta_{w_b^{|w_b|},b} -1$, with eigenvector 
$\bm{V_4} = (1,-1,-1,1)^t$, does not affect the RG transformation since 
$\bm{\mu}\cdot  \bm{V_4}  = 0$. Again, these properties generalize to 
$n$-letter substitutions with $|w_i|$ all odd. Here, isotropic aperiodicity in
general leads to contributions to $n-1$ additional scaling fields.
\end{itemize}
So far, the third case in the above classification had been set aside. 
Here, $|w_a|$ is odd and $|w_b|$ is even at the fixed point of 
$[\bm{M}_\varrho]_{\mbox{\footnotesize mod}\, 2}$. This makes things slightly 
more complicated since we cannot apply our pair substitution (\ref{pairsub}) here. 
Nevertheless, an exact renormalization scheme may be set up also in this case,
at least for two-letter substitution rules.
The main idea is not to construct a substitution rule for pairs of letters, 
but for all substrings of the chain with an even number of $a$'s and $b$'s and
of minimal length (that is they cannot be divided into smaller strings with the
same property). Obviously, $(aa)$ and $(bb)$ are examples for such minimal 
strings, a general string $s$ with length $2k \ge 4$ begins and ends with a pair 
of letters $ab$ or $ba$ with an arbitrary permutation of $k-1$ pairs $aa$ and
$bb$ in between: 
\begin{equation}
s = \left( {\left\{ \begin{matrix} ab\\ba \end{matrix} \right\} 
\left\{ \begin{matrix} aa\\bb\end{matrix} \right\}^{k-1} 
\left\{\begin{matrix}ab\\ba\end{matrix} \right\}}\right)   
\end{equation}
For a given substitution rule, the number of different minimal strings is 
always finite, hence a substitution rule on a finite ``alphabet of different
strings'' $s_i$ can always be found
\begin{equation}
\varrho_s : s_i \to w_{s_i} = \varrho(s_i^1)\varrho(s_i^2) \dots
\end{equation}
with substitution matrix
\begin{equation}
[\bm{M}_s]_{ij} = \#_{s_i}(w_{s_j}) 
\end{equation}
since $w_{s_i}$ may always be dissected into minimal strings.
For real space renormalization, in a first step, we contract the
strings by star multiplication of the corresponding $S$ transfer
matrices, this way assigning a single degree of freedom to each
string. After that we proceed the usual way, reversing the
substitution steps of the {\em string substitution} by decimation. Scaling fields
and renormalization group eigenvalues are again determined by the
action of the transpose of the substitution matrix $\bm{M}_s$ on a scaling vector
$\bm{\mu}$ with entries corresponding to the different strings, where, 
from the initial conditions,
\begin{equation}
\mu_{s_i} 
= \Delta_a \#_a(s_i) + \Delta_b \#_b(s_i) + r (\#_{ab} - \#_{ba})(s_i) \; .
\end{equation}
A more detailed analysis of the string substitution is given in the appendix,
with the following results. As in the above cases, the anisotropy parameters
transform with (\ref{anisotran}), leading to the Ising-like renormalization flow.
For two-letter substitution rules, the effect of isotropic aperiodicity is very
similar to the case 2 above. A finite $r$ does not lead to a renormalization
of the anisotropy parameters, the criticality condition is given by (\ref{crit2}).
In particular, this means that the aperiodic XX model is always critical. 
But as in case 2, isotropic aperiodicity introduces a new scaling 
field in the model, with RG eigenvalue $\lambda_s =0$, 
if $(\#_{ab} - \#_{ba})(w_b) = 0$, and $\lambda_s = (2\#_{b0} - \#_b)(w_b)$ 
otherwise. 
Again, the question arises whether these results generalize to the $n$-letter 
case. Note first that the method presented here can be applied only to a 
subclass of $n$-letter substitution rules. In general, a transformation 
into a string substitution with strings of even length will not be possible.
Within these limits (but probably also beyond) the transformation of the anisotropy
parameters $\Delta_{\alpha\beta}$ generalizes just as in cases 1 and 2. 
On the other hand, isotropic aperiodicity in general will alter the 
criticality condition and renormalize $\bm{\Delta}$, but will always also 
introduce new scaling fields.

\subsection{Critical scaling behaviour}
The determination of the critical scaling behaviour of the lowest fermionic
excitations may now be performed along the same lines as in the case of the
aperiodic Ising quantum chain \cite{HGB}. We will therefore only give a short 
account here. For cases 1 and 2, the RG 
transformations of the weights and fermion frequencies are to linear order 
in $\Lambda$
\begin{equation} \label{lamtran}
\tilde{\bm{\Lambda}}_+ = \bm{M}^+ \bm{\Lambda}_+ \;; \quad
\tilde{\bm{\Lambda}}_- = \bm{M}^- \bm{\Lambda}_- 
\end{equation}
where $\bm{\Lambda}_\pm = \Lambda (\kappa_{aa}^\pm, \kappa_{ab}^\pm,
\kappa_{ba}^\pm, \kappa_{bb}^\pm)^t$ and 
\begin{align}\label{wtran}
M^+_{\alpha\beta,\alpha'\beta'} &=  \sum_{k=1}^{|w_{\alpha\beta}|/2}
\delta_{w_{\alpha\beta}^{2k-1}w_{\alpha\beta}^{2k},\alpha'\beta'}
\left(\frac{\varepsilon_\alpha^y}{\varepsilon_{w_{\alpha\beta}^{2k-1}}^y}
\prod_{\ell = 1}^{k-1} \exp({\mu_{w_{\alpha\beta}^{2\ell-1} 
w_{\alpha\beta}^{2\ell}}})\right)^2
\\
M^-_{\alpha\beta,\alpha'\beta'} &= \sum_{k=1}^{|w_{\alpha\beta}|/2}
\delta_{w_{\alpha\beta}^{2k-1}w_{\alpha\beta}^{2k},\alpha'\beta'}
\left(\prod_{\ell = k+1}^{|w_{\alpha\beta}|/2} 
\exp({\mu_{w_{\alpha\beta}^{2\ell-1} w_{\alpha\beta}^{2\ell}}})\right)^2\;.
\end{align}
For the derivation of (\ref{wtran}), the special form of the substitution rule
(\ref{subs}) is used. At the Onsager or marginal fixed points, a similarity 
transformation of $\bm{M}^+$ yields the more symmetric form \cite{HGB}
\begin{equation}
M^\pm_{\alpha\beta,\alpha'\beta'} = \exp(\mp 2\mu_{\alpha\beta}) 
\sum_{k=1}^{|w_{\alpha\beta}|/2}
\delta_{w_{\alpha\beta}^{2k-1}w_{\alpha\beta}^{2k},\alpha'\beta'}
\prod_{\ell = 1}^{k} \exp({\pm 2\mu_{w_{\alpha\beta}^{2\ell-1} 
w_{\alpha\beta}^{2\ell}}})
\end{equation}
In this form, the matrix elements are functions of the reduced couplings alone.
The transformations in case 3 are analogous. The vectors of the weights
and fermion frequencies converge under iteration of (\ref{lamtran}) to the 
Perron Frobenius eigenvectors of $\bm{M}^\pm$ and the scaling behaviour of 
the lowest fermionic excitations ($q \ll N$) is then given by the
normalization condition of $\bm{\Lambda}_\pm$ as \cite{HGB}
\begin{equation}
\Lambda_q \sim \left(\frac{q}{N}\right)^z \;;
\quad z = \frac{\ln(\lambda_{M+} \lambda_{M-})}{2\ln\lambda_+}
\end{equation}
where $\lambda_{M\pm}$ are the largest eigenvalues of $\bm{M}^\pm$.
For irrelevant aperiodic modulations, we obtain
\begin{gather}
\bm{M}^+ = \bm{M}^- = \bm{M}_2^t
\\
\lambda_{M+} = \lambda_{M-} = \lambda_+
\end{gather}
and hence $z=1$. Since the fixed-point values of weights and couplings
are independent of $r$ and $\bm{\Delta}$ and thus the same as for the 
uniform chain we may also conclude that the low energy excitations are
equally spaced, in accordance with the predictions of {\em conformal
invariance}. For the related problem of the Ising quantum chain, this 
has been observed numerically before (e.g.~\cite{GB}).

Near the marginal fixed points, the coupling constants and thus the eigenvalues
$\lambda_{M\pm}$ take non-trivial values. For aperiodic anisotropy ($r=0$), the
RG equations may exactly be reduced to the corresponding ones of the 
Ising quantum chain in a constant transverse magnetic field ($h=1$), 
with coupling constants $\varepsilon_{a,b}^2 = \exp(\Delta_{a,b})$. 
Indeed, numerical observations show that not only the scaling
exponents, but the entire low energy spectra are identical up to a
common factor (altered fermion velocity). Note that this
correspondence does of course not extend to the entire (high energy) spectrum.
The scaling exponent $z$ has been calculated explicitly for 
arbitrary two-letter substitutions in  \cite{HGB}. 
After a proper renormalization of $\bm{\Delta}$, also the 
exponents of marginal substitutions of case 1 are found easily through this 
correspondence. 
For substitution rules of the second case, isotropic and anisotropic aperiodic 
modulation may be independently marginal. Generically, this leads to
a scaling exponent $1 < z < \infty$ depending on as many parameters as 
marginal scaling fields are present in the problem. Especially in the 
$n$-letter case, however, there are also exceptional substitutions, where 
marginal RG eigenvalues are in fact {\em marginally irrelevant}, leading to 
$z=1$, or also lead to a {\em marginally relevant} scaling behaviour with 
$z = \infty$, see \cite{Diss} for a discussion. 

For relevant aperiodic modulations, the reduced coupling constants do not tend 
to a finite limit, but (generically) grow with the second largest 
eigenvalue $\lambda_2$ of $\bm{M}_2$. As a consequence, $\lambda_{M\pm}$ 
finally scales like $\lambda_{M\pm} \sim \exp(c \lambda_2^n)$ in the $n$th 
renormalization step, resulting in a scaling behaviour of the lowest gaps as
(see \cite{HGB} for a more detailed discussion in the Ising case)
\begin{equation}
\Lambda_q \sim \exp(-c (N/q)^{\omega_2}) 
\end{equation}
with $\omega_2$ defined in (\ref{omega2}). 
Again, the same scaling behaviour, with the wandering exponent
$\omega_\varrho$ of the sequence of couplings directly,  had been found 
for the Ising quantum chains \cite{L1}.
Note that, contrary to {\em random disorder}, for aperiodic
sequences in general $\omega_\varrho \neq \omega_2$.

In general, the RG flow to strong couplings may lead to rather unusual 
critical properties, where typical and mean values of various exponents
(e.g.~the correlation length critical exponent of the spin chain)
no longer coincide \cite{Fisher, IKR}, see also the discussion below. 
The scaling exponent $z$ to be calculated here is, however, not effected 
by ``untypical events'' of this type.

\section{Thermodynamical properties}

In this section the consequences of the scaling behaviour of the fermionic
spectrum to the thermodynamics of the spin chain are discussed. This may be
done in analogy to the analysis of the Fibonacci-XX-chain in \cite{LN}.

The critical scaling of the low-energy spectrum directly implies the scaling
form of the integrated density of states in the thermodynamical limit as \cite{LN,HGB}
\begin{equation} \label{h1}
H(\Lambda) \sim \Lambda^{1/z} g(\ln \Lambda/ \ln \lambda_+)\;; \quad \Lambda \to 0 
\end{equation} 
($g$ is a function with unit period) for marginal or irrelevant aperiodicity and
\begin{equation} \label{h2}
H(\Lambda) \sim (\ln |\Lambda|)^{-1/\omega_2} \;; \quad \Lambda \to 0
\end{equation} 
in the relevant case.
The free energy (per spin) of the XY- chain at finite temperature $1/\beta$ is 
given by an integral transform of the fermionic IDOS as \cite{Lieb,LN}
\begin{align}
\beta f &=  - \frac{1}{N}\sum_{q}\ln\left(1+\exp[\beta \Lambda_q]\right)
\\
&= - \int \text{d}H(\Lambda) \ln(1+\exp[\beta \Lambda]) \; .           
\end{align}
Now, the specific heat is given by 
\begin{equation}
C_v = \beta^2 \frac{\partial^2}{\partial \beta^2} \left[-\beta f \right]
= \frac{\beta^2}{4} \int \text{d}H(\Lambda) \frac{\Lambda^2}{\cosh^2(\beta \Lambda/2)} \;.
\end{equation}
At low temperature, this expression is dominated by the small $\Lambda$ region and
the $T\to 0$ scaling behaviour of $C_v$ is completely determined through the 
critical scaling of the fermionic spectrum 
\begin{equation}
C_v \sim T^{1/z} G(\ln T/\ln \lambda_+)  \;; \quad C_v \sim 1/(\ln T)^{1+1/\omega_2}
\end{equation}
for marginal (irrelevant) and relevant aperiodicity respectively,
$G$ is again a periodic function with unit period.
Similarly, the susceptibility at vanishing field in $z$-direction 
may be derived to leading order as (with $\Lambda(h) = \Lambda(h=0) 
+ h\cdot r(\Lambda)$, $r$ bounded)
\begin{equation}
\chi_z = - \frac{\partial^2 f(h)}{\partial h^2}\bigg|_{h=0} \sim
\frac{\beta}{4} \int  \frac{\text{d}H(\Lambda)}{\cosh^2(\beta \Lambda/2)} \;. 
\end{equation}
and
\begin{equation}
\chi_z \sim T^{1/z-1} G'(\ln T/\ln \lambda_+) \quad \text{resp.} \quad \chi_z \sim 
1/(T [\ln T]^{1/\omega_2})
\end{equation}
Thus the susceptibility diverges for any marginal or relevant aperiodic 
perturbation. Note that for $\omega_2 = 1/2$, which is the mean fluctuation 
exponent for uncorrelated random disorder, these expressions coincide with 
the scaling behaviour of the random chain \cite{Fisher}.

As stated above, the aperiodic XY spin chain is essentially equivalent to two 
decoupled tight binding models with aperiodic hopping. In this context, the scaling
exponent $z$ calculated above determines the {\em localization length} at half 
filling. It is well known that the one-dimensional tight-binding model with
random hopping exhibits a single delocalized state at the band center. This is 
also the case for any kind of aperiodic disorder (also due to random substitutions,
see below) fulfilling the criticality condition. Using the Thouless relation \cite{T} we 
obtain from (\ref{h1},\ref{h2}) a diverging localization length 
$\ell_\Lambda$ at $\Lambda=0$, like
\begin{align}
\ell_\Lambda &\sim \Lambda^{-1/z} \;;\qquad\qquad \Lambda \to 0 
\\
\ell_\Lambda &\sim |\ln \Lambda|^{-1+1/\omega} \;;\quad\,  \Lambda\to 0 
\end{align} 
for marginal (or irrelevant) and relevant perturbations respectively. A more detailed
analysis of the resulting state (extended or critical) is possible for particular
examples with the methods of \cite{KST}.  

\section{Examples}

\begin{itemize}
\item
The Thue-Morse chain, generated by
\begin{equation}
\varrho_{TM} : 
\left\{ \begin{array}{ccc} a&\to&ab\\ b&\to&ba \end{array} \right.
\end{equation}
belongs to case 1 and is critical whenever $\Delta_a = - \Delta_b$. 
The induced disorder is irrelevant, since $\#_{ab}(w_{ab}) = \#_{ba}(w_{ab})$.
\item
The period doubling chain, generated by 
\begin{equation}
\varrho_{pd} : 
\left\{ \begin{array}{ccc} a&\to&ab\\ b&\to&aa \end{array} \right.
\end{equation}
is an example of a substitution rule of case 1 with $|\lambda_-| = 1$.
The criticality condition reduces to $2\Delta_a + \Delta_b + r =0$; note in
particular that the $pd$ XX chain is not critical. The critical scaling 
exponent is well known from the Ising case \cite{IT}
\begin{equation}
z_{pd} = \frac{ \ln (2\cosh(\Delta_a/2))}{\ln 2}
\end{equation}
and is also for the XY- chain a function of only one variable. 
\item
The so-called precious mean chains \cite{Holz1} are generated by substitution rules
with a substitution matrix of the form
\begin{equation}
\bm{M}_{k} = \left(\begin{array}{@{\,}r@{\;\;}r@{\,}}
k&1\\1&0
\end{array} \right)
\end{equation}
with eigenvalues $\lambda_{k\pm} = (k\pm\sqrt{k^2+4})/2$.
According to the above classification, they all belong to case 2 and are
critical for $\lambda_{k+} \Delta_a = - \Delta_b$. While criticality only 
depends on the anisotropy parameters, the critical exponent depends solely on $r$.
Since $|\lambda_{k-}| < 1$, anisotropic precious-mean modulations are irrelevant. 
On the other hand, it is straightforward to check that $|\lambda_{xx}| =1$,
thus the isotropic aperiodicity is always {\em marginal}. 
For even $k$ ($k=2$ corresponds to the silver mean chain), we eliminate blocks 
corresponding to double substitution steps in the RG transformation and obtain
the scaling exponent
\begin{equation}
z_k = \frac{\ln \Theta_k}{\ln\lambda_{k+}}\;,\quad \Theta_k = \frac{1}{4} 
\left(k\rho + \sqrt{k^2\rho^2 +16}\right)
\end{equation}
where 
\begin{equation}\label{rho}
\rho = \exp(r/2) + \exp(-r/2) \; .
\end{equation}
For $k$ odd, $k = 2\ell-1$, in each RG step three substitution steps have to 
be reversed. We (finally) obtain the scaling exponent
\begin{equation}
\tilde{z}_k = \frac{\ln \Theta_\ell}{3\ln\lambda_{k+}} \;,\quad \Theta_\ell = 
\frac{1}{2}\left(P_\ell(r)\rho^2 + \sqrt{P^2_\ell(r)\rho^4 + 4}
\right)\,,
\end{equation}
where 
\begin{equation}
P_\ell(r) = \frac{\ell^2\sinh[\ell r] +(\ell-1)^2 \sinh[(\ell-1)r]}{\sinh[r]}
\end{equation}
and $\rho$ as defined above.
The first term of this series, with $k=\ell=P_1(r) \equiv 1$, 
corresponding to the Fibonacci chain, had already been obtained in 
\cite{LN} using the well-known properties of the Fibonacci trace map \cite{KKT}. 
Also for general precious mean chains, spectral scaling exponents may be 
calculated by trace maps due to the existence of invariants. This has been done 
in \cite{Holz2}. Note however that the scaling exponents found in \cite{Holz2} 
(given in terms of Chebyshev polynomials) do not simply translate to the above 
expressions since the transfer matrices of the aperiodic hopping problem do not 
have unit determinant. In contrast to the aperiodic potential problem 
\cite{Holz2}, 
the scaling exponents here behave differently for $k$ even or odd in the limit 
of weak incommensurability $k \to \infty$. For fixed ratio of the couplings $r$, 
aperiodicity becomes irrelevant for $k$ even ($\lim_{k\to\infty} z_k = 1$), but 
not for $k$ odd ($\tilde{z}_k \to \infty$).

The precious mean chains are just those quasicrystalline chains that result 
from the so-called cut-and-project formalism which the slope of the cut space 
given by $\lambda_{k+} = [0,\bar{k}]$. By the sucessive application of different 
precious mean substitutions, a much more general class of cut-and-project chains
may be generated. Indeed, since the eigenvector $\bm{v}_{xx}$ to the marginal
eigenvalue $\lambda_{xx}$ of $\bm{M}_2$ is independent of $k$, the marginal 
scaling property holds also for this more general class of chains. Quadratic 
irrationalities in particular, which are observed in real quasicrystalline 
matter, are given by periodic continuous fractions and lead to cut-and-project
chains that may be generated by a periodic application of precious mean 
substitutions. Thus also the scaling exponent $z$ can be calculated using the 
above method. 

Finally, we like to stress that the origins of the marginal scaling behaviour 
observed for the interface roughness of Fibonacci surfaces \cite{HeL,GL} and
the XY quantum chains on the other hand are independent. According to the 
Harris-Luck criterium, the former is connected to the fact that the unperturbed 
correlation length exponent which enters (\ref{HL}) is $\nu =1/2$ there and
the precious mean substitutions, being volume preserving, lead to the marginal 
wandering exponent $\omega_\varrho = -1$ \cite{I}. Substitution rules that lead to 
marginal scaling in only one of these situations are easily constructed; it is 
by chance that the precious mean chains fulfill both marginality conditions.

\item
Different types of the three-folding chain are defined by substitution rules
with substitution matrix
\begin{equation}
\bm{M}_{3f} = \left(\begin{array}{@{\,}r@{\;\;}r@{\,}}
2&1\\1&2
\end{array} \right)
\end{equation}
This is one of the simplest examples of a chain where both isotropical and
anisotropical aperiodicity are independently marginal (with exception of the 
special form $\varrho: a \to aba ; \, b\to bab$ which leads to a periodic 
chain and $z=1$).
At criticality ($\Delta_a = -\Delta_b\equiv \Delta$) we thus obtain 
continuously varying scaling exponent depending on two variables
\begin{equation}
z_{3f} = \frac{\ln (\rho^2+2\rho \cosh[\Delta]+1)}{2\ln 3}
\quad \text{for} \quad \varrho_{3f} :
\begin{array}{rcl}
a &\to& aab \\ b & \to & bab 
\end{array}
\end{equation}
and substitution rules that lead to {\em locally isomorphic} (or patch equivalent)
chains, and
\begin{equation}
\hat{z}_{3f} = \frac{\ln \Theta_{3f}}{\ln 3} 
\quad \text{for} \quad \hat{\varrho}_{3f} :
\begin{array}{rcl}
a &\to& aab \\ b & \to & bba 
\end{array}
\end{equation}
where $\Theta_{3f} =\rho\cosh(\Delta)+\sqrt{[2\sinh(r/2)\sinh(\Delta)]^2 +1}$.
For $r = 0$, these expressions reduce to the corresponding ones of the
Ising quantum chain \cite{IT}. Note also that $z_{3f} = \hat{z}_{3f}$
for pure isotropic aperiodicity.
\item
The substitution rule
\begin{equation}
\varrho : \left\{\begin{array}{ccl} a&\to&abb \\ b&\to&ababbb \end{array}\right.
\end{equation} 
belongs to case 3 of the classification above. A set of four strings,
\begin{equation}
\{(bb), (abab), (abba), (babbba)\}\;,
\end{equation}
is sufficient to define a string substitution with substitution matrix
\begin{equation} \label{stringsubs}
\bm{M}_{s} = \left( \begin{array}{@{\,}r@{\;\;}r@{\;\;}r@{\;\;}r@{\,}}
2&2&1&2\\2&1&0&1\\0&1&1&1\\0&1&2&3
\end{array} \right) \;.
\end{equation}
We have $\lambda_- =0$, hence aperiodic anisotropy is irrelevant, but since
$\lambda_s =2$ and $\omega_2 = \ln 2/ \ln 5 > 0$ isotropic
aperiodicity will be relevant. 
\end{itemize}

\section{Extensions to correlated random disorder}

Aperiodic order, as generated by substitution rules, represents a natural,
but non-trivial extension of crystalline and quasicrystalline order. 
Structures with this type of long-range order are certainly
physically reasonable (perhaps in contrast to hierarchical systems)  
but of course also show a number of quite special properties in
comparison to random systems. These include rescaling symmetries
and the {\em strong repetitivity} of local patches due to their 
self-similar structure. Moreover, the ordering is deterministic by
construction and leads to zero entropy density. 
However, as has been argued in \cite{L1,L2}, the thermodynamical
properties of (quantum) spin models should be unaffected by most of
these special properties, but depend only on the nature of the
fluctuations present in the system. In fact, also our RG formalism may
be applied to a more general class of models and in particular does
{\em not} depend on the exact self-similarity of the substitution chains. 
For simplicity, we concentrate on aperiodic anisotropy (respectively the Ising case).
Consider a chain of couplings chosen according to the following {\em
random substitution rule}
\begin{equation}
\varrho: a_i \to \text{perm}(w_i)
\end{equation}
where $\text{perm}(w_i)$ denotes a random permutation of the letters
in $w_i$. The class of chains generated this way is clearly neither
deterministic nor strongly repetitive, in fact it is, almost surely not
repetitive at all. What is more, its entropy density is
positive. Indeed, the nature of the fermionic spectrum is completely
changed by the introduction of randomness: whereas it is typically
purely {\em singular continuous} with a characteristic gap-structure
for substitution chains, all these gaps vanish in (numerical) spectra
of random-substitutions. As the only property that remains unchanged, the total
fluctuation of the mean (reduced) coupling still decomposes into a superposition
of a finite number of fluctuation modes, implicitly given through the
eigenvalues of the substitution matrix. This property is also the
essential ingredient for our RG procedure, which depends on the substitution
matrix rather than on the detailed form of the substitution
itself. Consequently, neither the RG flow nor the fixed point
structure are affected by introducing randomness into the substitution
rules. Note however that the scaling exponent $z$ in the marginal case
depends on $\varrho$ in more detail and we only obtain analytical upper
and lower bounds ($1 < z_1 \le z \le z_2 < \infty$) for random 
substitutions here \cite{Diss}. Numerical results show that $z$ may
indeed vary within this interval and does not converge to a well 
defined limit. 

In many respects, random substitution chains with relevant fluctuation modes 
(in particular those with wandering exponent $\omega_2 = 1/2$) behave
very similar to uncorrelated random chains. However, a characteristic 
difference is that for the latter only the asymptotic growth of the {\em mean}
fluctuations is controlled by the mean deviation exponent $\omega =
1/2$, while fluctuations of any order (up to $\omega = 1$) may be
present with a non-vanishing probability on every lengthscale. The
most significant consequence is the off-critical Griffiths phase
observed in random quantum chains \cite{G,Fisher}, but not in aperiodic models
\cite{IKR}. Also for random substitution chains with exponentially
many realizations, no Griffiths phase should be present since for any
non-critical values of the coupling constants there is a finite maximal 
size for ``locally critical'' patches.

\section{Discussion}
We extended an exact real space renormalization approach, originally formulated
for Ising quantum chains to aperiodic XY quantum chains. This way, relevance 
criteria for aperiodic modulations have been obtained analytically for a 
second class of models. As predicted by the Harris-Luck relevance criterion, 
the geometrical fluctuation exponent plays the key role in the 
determination of the critical behaviour. However, the fluctuation exponent
$\omega_2$ of the sequence of ratios of consecuting $\varepsilon^x$
and $\varepsilon^y$ couplings, which 
matters for the XY models, may differ from the wandering exponent 
$\omega_\varrho$ of the sequence of interactions directly, which had been the 
crucial quantity in the Ising case. As a consequence, the relevance of the same 
aperiodic ordering of the coupling constants in general will be different for 
Ising quantum chains and isotropic XX models. In particular, quasi-periodic 
disorder, generated by substitution rules compatible with the cut-and-project 
formalism, is irrelevant for Ising quantum chains and most other Ising spin
systems, but marginal for XX or XY chains. On the other hand, aperiodic XY 
anisotropy leads to identical fluctuations and the same critical behaviour 
as in the Ising case. 

The analysis of the RG fixed point structure and renormalization flows in 
particular indicates that there is no discrimination between weak and strong 
aperiodic disorder in these models. The validity of the perturbative Harris-Luck
criterion is thus extended to the case of strong modulations. 
Open questions remain mainly for relevant aperiodic disorder. Here the RG flows
to the strong coupling limit and the critical scaling behaviour of several ensemble
averaged quantities is dominated by rare events. A comparison of the resulting
``aperiodic ground states'' to the so-called ``random singlet phase'' postulated
for uncorrelated random chains \cite{Fisher} would be of interest. For the Ising
quantum chains, a first step into that direction has been done in \cite{IKR}. 
We have shown that the RG approach may be also applied to random substitutions
and does not rely too much on special properties of deterministic aperiodic
systems. Let us remark that -- especially in the Ising case -- the structure
of the RG is rather simple and an extension to ensembles of uncorrelated 
random chains should be possible somehow. The crucial question is whether
the atypical means of quantities like the critical correlation function will be
accessible within a RG scheme which works in the fermion representation.
  
The renormalization approach leads to an exact determination of the scaling   
exponent $z$ of the mass gap for arbitrary two-letter substitution rules and
quite general classes of $n$-letter substitutions. The critical exponents 
connected with the scaling of the spectrum at $\Lambda = 0$ may be calculated 
exactly, like the zero temperature specific heat, the susceptibility in a vanishing 
magnetic field in $z$-direction, or the localization length of the aperiodic
hopping model at half filling. We have given a number of quantitative results as 
examples, mainly for marginal aperiodicity.

The exact results obtained for the aperiodic XY models should be of use for
the analysis (analytical and numerical) of more complicated aperiodic models, 
like Heisenberg and XYZ spin chains. 
As for analytical results, a natural step would be to introduce a 
non-vanishing transverse magnetic field in $z$ direction. However, although 
this still leads to a free fermion model, a treatment by exact renormalization 
as shown here at zero field does not seem to be a simple problem.

\section{Acknowledgments}

The author wishes to thank Michael Baake for useful comments on the
manuscript. It is a pleasure to thank Prof.~Ralph v.~Baltz and the
``Institut f{\"u}r Theorie der kondensierten Materie'',
Karlsruhe University, for hospitality, where part of this work was
done. This work was supported by Deutsche Forschungsgemeinschaft and
Studienstiftung des Deutschen Volkes.

\setcounter{section}{0}
\renewcommand{\thesection}{Appendix \Alph{section}:}
\renewcommand{\thesubsection}{\Alph{section}.\arabic{subsection}}
\renewcommand{\theequation}{\Alph{section}.\arabic{equation}}

\section{Spectra of string substitutions}
\setcounter{equation}{0}

We are interested in the action of the transpose string substitution
matrix $\bm{M}_s^t$ on the vector $\bm{\mu}$ with
\begin{equation}
\mu_{s_i} 
= \Delta_a \#_a(s_i) + \Delta_b \#_b(s_i) + r (\#_{ab} - \#_{ba})(s_i) \; .
\end{equation}
For $\tilde{\bm{\mu}} \equiv \bm{M}_s^t \bm{\mu}$ we find
\begin{equation}
\tilde{\mu}_{s_i} = \Delta_a \#_a(w_{s_i}) + \Delta_b \#_b(w_{s_i}) +
r(\#_{ab}-\#_{ba})(w_{s_i}) \;.
\end{equation}
Using
\begin{align}
\#_a(w_{s_i}) &= \#_a(w_a)\#_a(s_i) + \#_a(w_b)\#_b(s_i) \\
\#_b(w_{s_i}) &= \#_b(w_a)\#_a(s_i) + \#_b(w_b)\#_b(s_i)
\end{align}
we obtain the transformation rule (\ref{anisotran}) of the anisotropy parameters.
\\ 
On the other hand, since $|w_a|$ is odd and $|w_b|$ even, we may write
\begin{equation}
(\#_{ab} - \#_{ba})(w_{s_i}) = (\#_{ab} - \#_{ba})(w_b) \cdot (2\#_{b0} - \#_{b})
(s_i)
\end{equation}
where $\#_{b0}(s_i)$ gives the number of $b$'s in $s_i = s_{i1}bs_{i2}$ with
$\#_a(s_{i1})$ even. Thus isotropic aperiodicity is clearly irrelevant if
$(\#_{ab} - \#_{ba})(w_b) =0$. Otherwise, consider now the action of $\bm{M}_s^t$
on $\tilde{\bm{\mu}}$. For $\bm{\Delta} = 0$ we obtain
\begin{align}
\left[\bm{M}_s^t \tilde{\bm{\mu}}\right]_{s_i} &= 
r (\#_{ab} - \#_{ba})(w_b) \cdot (2\#_{b0} - \#_{b})(w_{s_i})
\\
&= r (\#_{ab} - \#_{ba})(w_b) \cdot (2\#_{b0} - \#_{b})(w_b)\cdot 
(2\#_{b0} - \#_{b})(s_i)
\end{align}
and recognize $\lambda_s \equiv (2\#_{b0} - \#_{b})(w_b)$ as the desired 
RG eigenvalue. Since $|w_b|$ is even, so is $\lambda_s$, isotropic aperiodic
modulation is thus either irrelevant or relevant, but never marginal in this 
case.


\begin{thebibliography}{99}
\bibitem{AS}
J.~Ashraff and R.~Stinchcombe,
       Exact decimation approach to the Green's functions of the
       Fibonacci-chain quasicrystal,
       {\it Phys.~Rev.\/} {\bf B37} (1988) 5723--29.
\bibitem{BGB}
M.~Baake, U.~Grimm and R.\thinspace J.~Baxter,
       A critical Ising model on the labyrinth,
       {\it Int.~J.~Mod.~Phys.\/} {\bf B8} (1994) 3579--600.
\bibitem{Fisher}
D.~Fisher,
       Random antiferromagnetic quantum spin chains,
       {\it Phys.~Rev.\/} {\bf B50} (1994) 3799--821.
\bibitem{GLO}
C.~Godr\'eche, J.\thinspace M.~Luck and H.~Orland,
       Magnetic Phase Structure on the Penrose Lattice,
       {\it J.~Stat.~Phys.\/} {\bf 45} (1986) 777--800.
\bibitem{G}
R.~Griffiths,
       Nonanalytic behavior above the critical point in a random Ising ferromagnet,
       {\it Phys.~Rev.~Lett.\/}{\bf 23} (1969) 17--19.
\bibitem{GB}
U.~Grimm and M.~Baake,
       Nonperiodic Ising Quantum Chains and Conformal Invariance
       {\it J.~Stat.~Phys.\/} {\bf 74} (1993) 1233--45.
\bibitem{GB2}
U.~Grimm and M.~Baake,
       Aperiodic Ising Models, in
       {\it The Mathematics of Long-Range Aperiodic Order},
       ed.~R.~V.~Moody, Kluwer, Dordrecht (1997), pp.~199--237. 
\bibitem{GL}
A.~Garg and D.~Levine,
       Faceting and Roughening in Quasicrystals
       {\it Phys.~Rev.~Lett.\/} {\bf 59} (1987) 1683--86.
\bibitem{Harris}
A.\thinspace B.~Harris,
       Effect of random defects on the critical behaviour of Ising models,
       {\it J.\ Phys.\/} {\bf C7} (1974) 1671--92.
\bibitem{HeL}
C.\thinspace L.~Henley and R.~Lipowsky,
       Interface Roughening in Two-Dimensional Quasicrystals
       {\it Phys.~Rev.~Lett.\/} {\bf 59} (1987) 1679--82.
\bibitem{HGB}
J.~Hermisson, U.~Grimm and M.~Baake,
       Aperiodic Ising quantum chains,
       {\it J.~Phys.\/} {\bf A30} (1997) 7315--35.
\bibitem{Diss}
       J.~Hermisson, PhD-thesis, in preparation.
\bibitem{Holz1}
M.~Holzer,
       Three classes of one-dimensional, two-tile Penrose tilings and the
       Fibonacci Kronig-Penny model as a generic case,
       {\it Phys.~Rev.\/} {\bf B38} (1988) 1709--20.
\bibitem{Holz2}
M.~Holzer,
       Nonlinear dynamics of localization in a class of one-dimensional
       quasicrystals
       {\it Phys.~Rev.\/} {\bf B38} (1988) 5756--59.
\bibitem{IKR}
F.~Igl\'{o}i, D.~Karevski and H.~Rieger,
       Random and aperiodic quantum spin chains: A comparative study,         
       {\it Eur.~Phys.~J.} {\bf B1} (1998) 513--17.
\bibitem{KKT}
M.~Kohomoto, L.~Kadanoff and C.~Tang,
       Localization Problem in One Dimension: Mapping and Escape,
       {\it Phys.~Rev.~Lett.\/} {\bf 50} (1983) 1870--72.
\bibitem{KST}
M.~Kohomoto, B.~Sutherland and C.~Tang,
       Critical wave functions and a Cantor-set spectrum of a 
       one-dimensional quasicrystal model,
       {\it Phys.~Rev.\/} {\bf B35} (1986) 1020--33.
\bibitem{I}
F.~Igl\'{o}i,
       Critical behaviour in aperiodic systems,
       {\it J.~Phys.\/} {\bf A26} (1993) L703--9.
\bibitem{IT2}
F.~Igl\'{o}i and L.~Turban,
       Relevant Aperiodic Modulation in the $2d$ Ising Model,
       {\it Europhys.~Lett.\/} {\bf 27} (1994) 91--96.
\bibitem{IT}
F.~Igl\'{o}i and L.~Turban,
       Common trends in the critical behavior of the Ising and
       directed walk models, 
       {\it Phys.~Rev.~Lett.\/} {\bf 77} (1996) 1206--9.
\bibitem{Lieb}
E.\thinspace H.~Lieb, T.~Schultz and D.~Mattis,
       Two Soluble Models of an Antiferromagnetic Chain,
       {\it Ann.~Phys.~(NY)} {\bf 16} (1961) 407--66.
\bibitem{LN}
J.\thinspace M.~Luck and Th.~Nieuwenhuizen,
       A Soluble Quasi-Crystalline Magnetic Model: the XY Quantum Spin
       Chain,
       {\it Europys.~Lett.\/} {\bf 2} (1986) 257--66.
\bibitem{L1}
J.\thinspace M.~Luck, 
       Critical Behavior of the Aperiodic Quantum Ising
       Chain in a Transverse Magnetic Field,
       {\it J.\ Stat.\ Phys.\/} {\bf 72} (1993) 417--58.
\bibitem{L2}
J.\thinspace M.~Luck,
       A classification of critical phenomena on quasi-crystals
       and other aperiodic structures,
       {\it Europhys.\ Lett.\/} {\bf 24} (1993) 359--64.
\bibitem{ON1}
Y.~Okabe and K.~Niizeki,
       {J.~Phys.~Soc.~Japan} {\bf 57} (1988) 1536.
\bibitem{ON2}
Y.~Okabe and K.~Niizeki,
       Ising model on an icosahedral quasilattice,
       {J.~Phys.\/} {\bf A23} (1990) L733--38.
\bibitem{Pfeuty}
P.~Pfeuty,
       An exact result for the 1D random Ising model in a transverse
       field, {\it Phys.~Lett.\/} {\bf A72} (1979) 245--46. 
\bibitem{Q}
M.~Queffelec,
       {\it Substitution Dynamical Systems}, (Lecture Notes in Mathematics, 
        vol.~1294) (Springer, Berlin 1987).
\bibitem{RED}
R.~Redheffer,
       Difference equations and functional equations in transmission line theory,
       in: {\it Modern Mathematics for Engineers}, ed.~E.~Beckenbach 
       (McGraw-Hill, New York 1961) pp 282--337.
\bibitem{S}
H.~Simon,
       {\it Ferromagnetische Isingmodelle auf aperiodischen Strukturen},
       PhD thesis, Universit{\"a}t T{\"u}bingen,
       Dissertations Druck Darmstadt (1997).
\bibitem{Smith}
E.~Smith,
       One-dimensional X-Y model with random coupling
       constants. I.~Thermodynamics, 
       {\it J.~Phys.\/} {\bf C3} (1970) 1419--32.
\bibitem{SJR}
S.~Sorensen, M.Jari\'{c} and M.~Ronchetti,
       Ising model on Penrose lattices: Boundary conditions,
       {\it Phys.~Rev.\/} {\bf B44} 9271--82.
\bibitem{T}
D.~J.~Thouless,
       A relation between the density of states and range of localization for
       one dimensional random systems,
       {\it J.~Phys.\/} {\bf C5} (1971) 77--81.       
\bibitem{TBB}
L.~Turban, P.-E.~Berche and B.~Berche,
       Surface magnetization of aperiodic Ising systems: a comparative study of
       bond and site problems,
       {J.~Phys.\/} {\bf A27} (1994) 6349--66.
\end{thebibliography}
\end{document}